\documentclass[12pt]{article}
\usepackage{amsfonts}
\usepackage{amsmath}
\usepackage{epsfig}
\usepackage{graphicx}

\begin{document}

\title{\textbf{Portfolios and the market geometry}}
\author{Samuel Eleut\'{e}rio\thanks{%
Instituto Superior T\'{e}cnico, Av. Rovisco Pais 1049-001 Lisboa, Portugal,
sme@ist.utl.pt}, Tanya Ara\'{u}jo\thanks{
ISEG, Technical University of Lisbon, Rua do Quelhas, 6 1200-781 Lisboa
Portugal, tanya@iseg.utl.pt} and R. Vilela Mendes\thanks{%
Centro de Matem\'{a}tica e Aplica\c{c}\~{o}es Fundamentais, Av. Gama Pinto
2, 1649-003 Lisboa, Portugal, vilela@cii.fc.ul.pt}}
\date{ }
\maketitle

\begin{abstract}
A geometric analysis of the time series of returns has been performed in the
past and it implied that the most of the systematic information of the
market is contained in a space of small dimension. Here we have explored
subspaces of this space to find out the relative performance of portfolios
formed from the companies that have the largest projections in each one of
the subspaces. It was found that the best performance portfolios are
associated to some of the small eigenvalue subspaces and not to the dominant
directions in the distances matrix. This occurs in such a systematic fashion
over an extended period (1990-2008) that it may not be a statistical
accident.
\end{abstract}


\section{Introduction}

Correlations in return fluctuations of market securities play an important
role in the analysis of market structure \cite{Mantegna1}, forecasting and
portfolio theory. The quantity%
\begin{equation}
d_{kl}=\sqrt{2\left( 1-C_{kl}\right) }  \label{I.1}
\end{equation}%
where $C_{kl}$ is the correlation coefficient of two (return) time series $%
\overrightarrow{r}(k)$ and $\overrightarrow{r}(l)$ 
\begin{equation}
C_{kl}=\frac{\left\langle \overrightarrow{r}(k)\overrightarrow{r}%
(l)\right\rangle -\left\langle \overrightarrow{r}(k)\right\rangle
\left\langle \overrightarrow{r}(l)\right\rangle }{\sqrt{\left( \left\langle 
\overrightarrow{r}^{2}(k)\right\rangle -\left\langle \overrightarrow{r}%
(k)\right\rangle ^{2}\right) \left( \left\langle \overrightarrow{r}%
^{2}(l)\right\rangle -\left\langle \overrightarrow{r}(l)\right\rangle
^{2}\right) }}  \label{I.2}
\end{equation}%
($r_{t}(k)=\log (p_{t}(k))-\log (p_{t-1}(k))$), has been shown \cite%
{Mantegna2} to satisfy all the metric axioms. Hence it may be used as a
basis to develop a geometrical analysis of the market structure. Such an
analysis has been performed in Ref.\cite{VM-PhysA}. Given a matrix of
distances, obtained from (\ref{I.1}), for a set of $N$ time series in a time
window, one obtains coordinates in $R^{N-1}$ compatible with these
distances. For that time window the returns of the companies are now
represented by a set $\left\{ x_{i}\right\} $ of points in $R^{N-1}$.
Assigning to each point a mass proportional to the market capitalization,
the center of mass $\overrightarrow{R}$ and the center of mass coordinates $%
\overrightarrow{y}(k)=\overrightarrow{x}(k)-\overrightarrow{R}$ are
obtained. Then the tensor%
\begin{equation}
T_{ij}=\Sigma _{k}\overrightarrow{y_{i}}(k)\overrightarrow{y}_{j}(k)
\label{I.3}
\end{equation}%
is diagonalized to obtain the set of eigenvalues and normalized eigenvectors
\{$\lambda _{i},\overrightarrow{e}_{i}$\}. The eigenvectors $\overrightarrow{%
e}_{i}$ define the characteristic directions of the set of stocks. The same
analysis is performed for random and time permuted data and the relative
behavior of the eigenvalues is compared. Having carried out this analysis
for the companies in the Dow Jones and SP500 indexes \cite{VM-PhysA} \cite%
{Araujo-Louca1} the following conclusions were obtained:

1 - The eigenvalues decrease very fast and soon become indistinguishable
from those obtained from random data. It means that the systematic
information related to the market structure is contained in a reduced
subspace of low dimension. From the extensive amount of data that was
analyzed (ranging from 70 to 424 stocks) one concludes that the dimension $d$
of this subspace is at most six.

2 - The characteristic directions in the reduced subspace do not in general
correspond to the traditional industrial sectors, mixing companies of many
different sectors, thus showing the interlocked nature of the market. The
characteristic directions provide a natural basis for a model of market
factors.

3 - Carrying out the geometric analysis over many different periods, some
noticeable differences were found between business-as-usual and crisis
periods. During market crisis there is a contraction of volume in the
reduced space. It corresponds to a greater synchronization of the market
fluctuations. In addition, whereas the geometric "market cloud" of points in
business-as-usual periods looks like a smooth ellipsoid, during some crisis
it displays distortions, which may be detected by computing higher moments
of the distribution. Whether these distortions appear sufficiently ahead of
the crisis to act as precursors is still an open question.

The question that is addressed in this paper is whether the geometric market
structure and the characteristic dimensions have any bearing on the
construction of portfolios. From the geometric analysis we have seen that
part of the correlations is indistinguishable from those in random data, the
market systematic structure being carried by a smaller $d-$dimensional
subspace. On the other end the characteristic directions are uncorrelated to
each other. Therefore, when forming portfolios that mimic these directions
one is exploring the systematic components of the market. There is no
obvious relation between the geometrical status of the market directions and
the return performance of its associated portfolio. To explore this issue
the following experimental approach was used:

A specific time interval, herein called \textit{the past,} is used to
construct the effective dimensions of the market. For each direction,
portfolios are formed with the companies that have projections along this
direction above a threshold. They are called \textit{dominant} for that
direction. Then the behavior of these portfolios is followed for a later
time interval, called \textit{the future}. Afterwards a new dimension
analysis is performed using the data of the period called "the future" and
the portfolio is adjusted accordingly. Portfolios were also formed mixing
dominant companies in several directions. Carrying out this analysis for the
data of 20 years some surprising results were found. The portfolios
corresponding to the largest eigenvalue directions tend to perform poorly,
whereas it is some of the smaller eigenvalues portfolios that perform
better. Whether this is an accident or there is some deep reason it is an
open question. In any case if it is an accident, it is an accident that
consistently occurs over very many years.

In the final section we place our portfolios in the Markowitz plane. It
turns out that the better performing ones are close to a frontier portfolio.

\section{Portfolios and characteristic market directions: An empirical study}

We have analyzed a set of 319 NYSE stocks for a time period from 1989 to
2008. Once the characteristic market directions (for a time interval called 
\textit{the past}) and the reduced $6-$dimensional subspace are identified,
the first step consists in determining the amount of each stock to be
included in each one the directional-portfolios. Once this is done, the
performance of the portfolios is followed for a time period called \textit{%
the future }and is compared with the performance of the Dow Jones and SP500
indexes. Afterwards the data of \textit{the future} is used to redo the
geometrical analysis and new directional-portfolios (using the accumulated
capital of the previous ones) are formed which are then followed for an
equal period. etc. For the results presented in this paper past and future
are six months.

To obtain the contribution of each stock ($i$) to a $d-$dimensional
subspace\ $\Omega $, we compute the ratio between the projection of the
stock to that subspace and the projection to the whole market space $%
(R^{N-1})$.

\begin{equation*}
f(i,\Omega )={\frac{\sqrt{\sum_{\alpha \in \Omega }x(i,\alpha )^{2}}}{\sqrt{%
\sum_{\alpha \in R^{N-1}}x(i,\alpha )^{2}}}}
\end{equation*}

The inclusion of the stock ($i$) in the $d$-dimensional portfolio depends on
the value of $f(i,\Omega )$, which is required to be greater than an
appropriate threshold. The weight of stock $i$ in the portfolio is
proportional to $f(i,\Omega )$. At time zero the portfolio is normalized
using the value of an index (Dow Jones or SP500) at that time.

On the final day of each time period ($6$ months) the ratio $D_{\Omega }$ of
the portfolio value $P_{t}(\Omega )$ to the reference index value on that
day $P_{t}\left( Idx\right) $ is computed:

\begin{equation*}
D_{\Omega }=\frac{P_{t}(\Omega )}{P_{t}\left( Idx\right) }
\end{equation*}

\subsection{One-dimensional portfolios}

\begin{figure}[tbp]
\centering
\includegraphics[width=10cm]{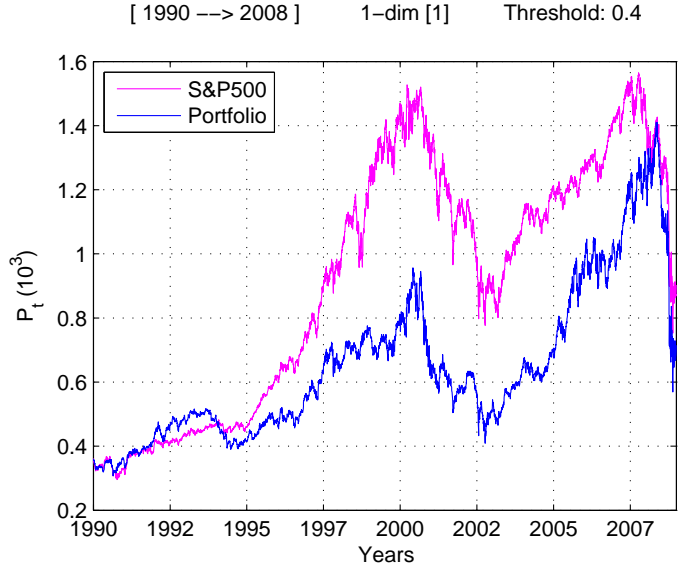}
\caption{Direction-1 portfolio (with threshold 0.4) compared with S\&P500}
\label{dim1_0.4}
\end{figure}

\begin{figure}[tbp]
\centering
\includegraphics[width=10cm]{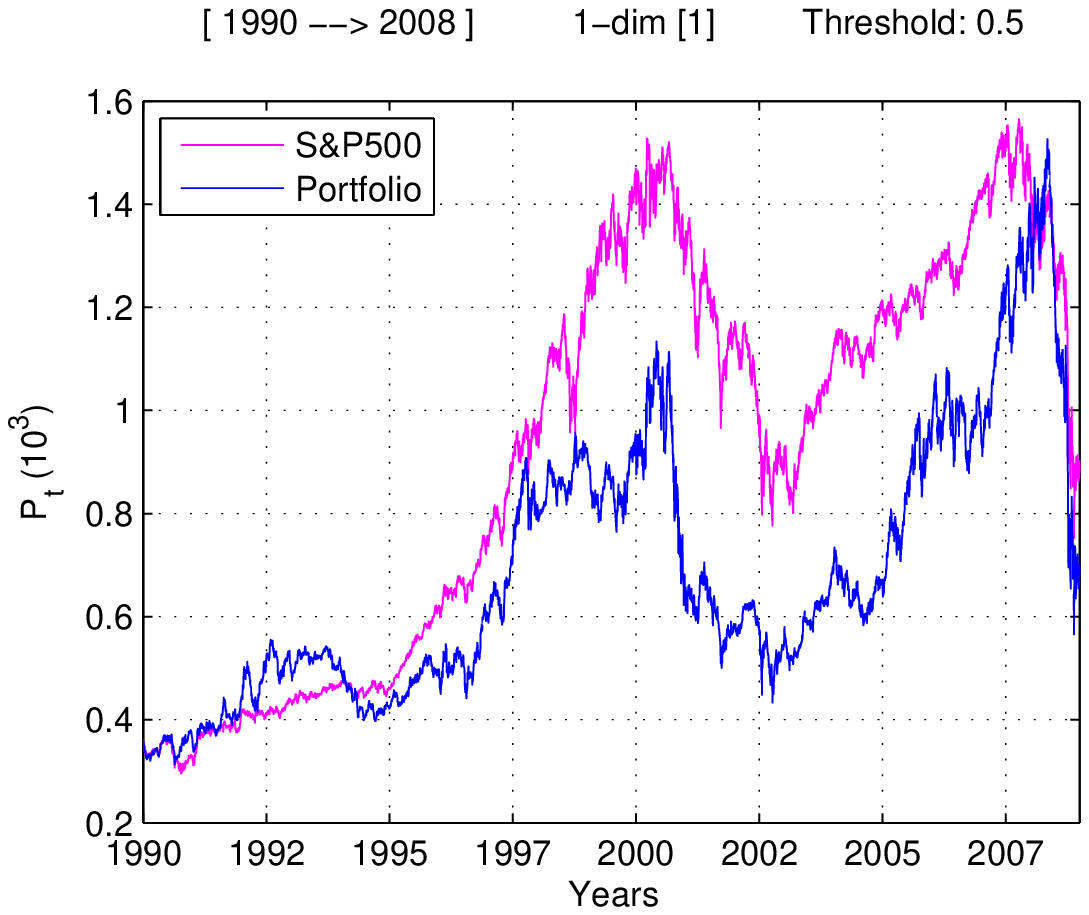}
\caption{Direction-1 portfolio (with threshold 0.5) compared with S\&P500}
\label{dim1_0.5}
\end{figure}

\begin{figure}[tbp]
\centering
\includegraphics[width=10cm]{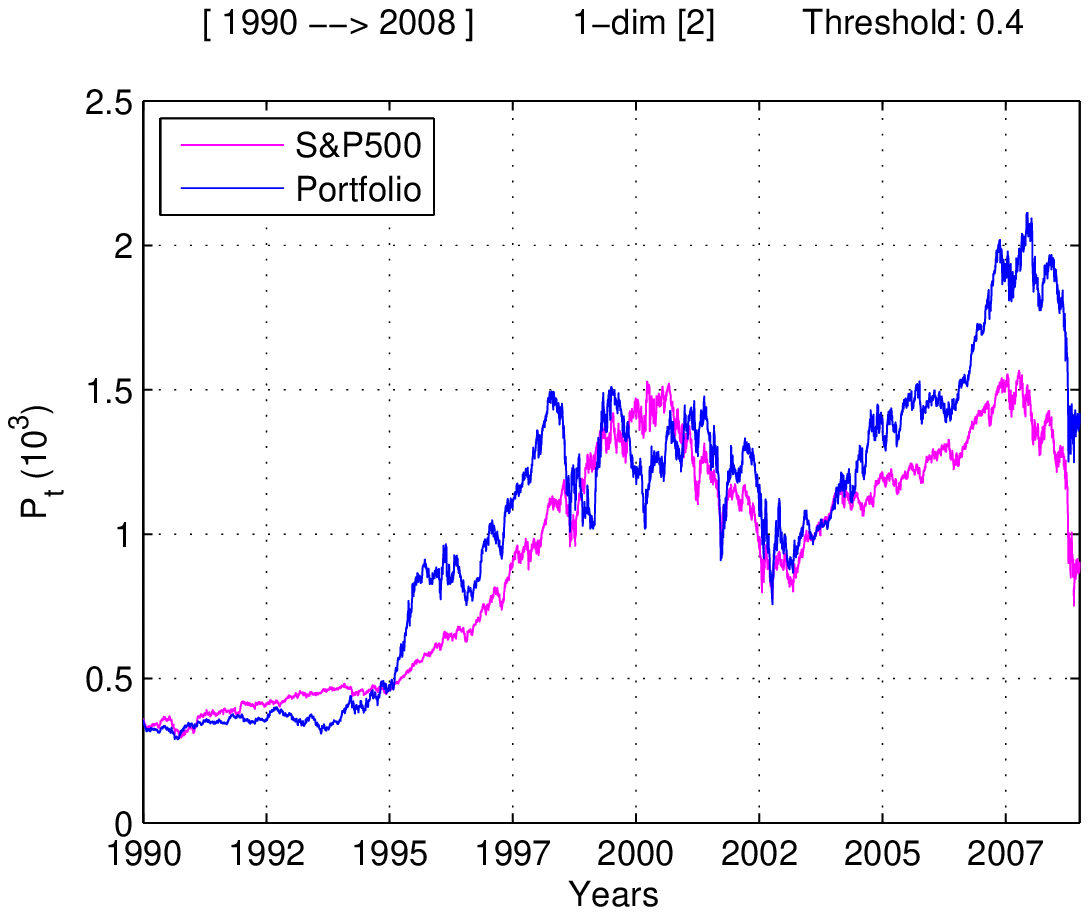}
\caption{Direction-2 portfolio (with threshold 0.4) compared with S\&P500}
\label{dim2}
\end{figure}

\begin{figure}[tbp]
\centering
\includegraphics[width=10cm]{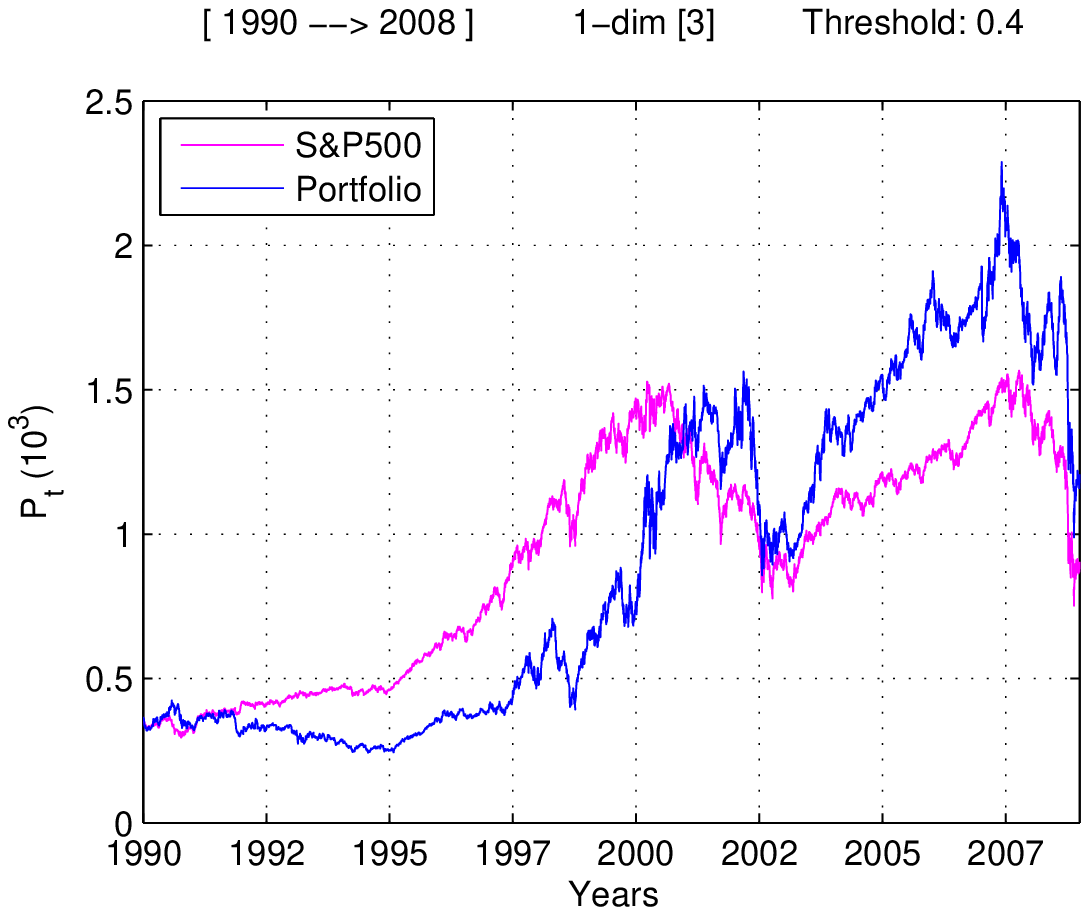}
\caption{Direction-3 portfolio (with threshold 0.4) compared with S\&P500}
\label{dim3}
\end{figure}

\begin{figure}[tbp]
\centering
\includegraphics[width=10cm]{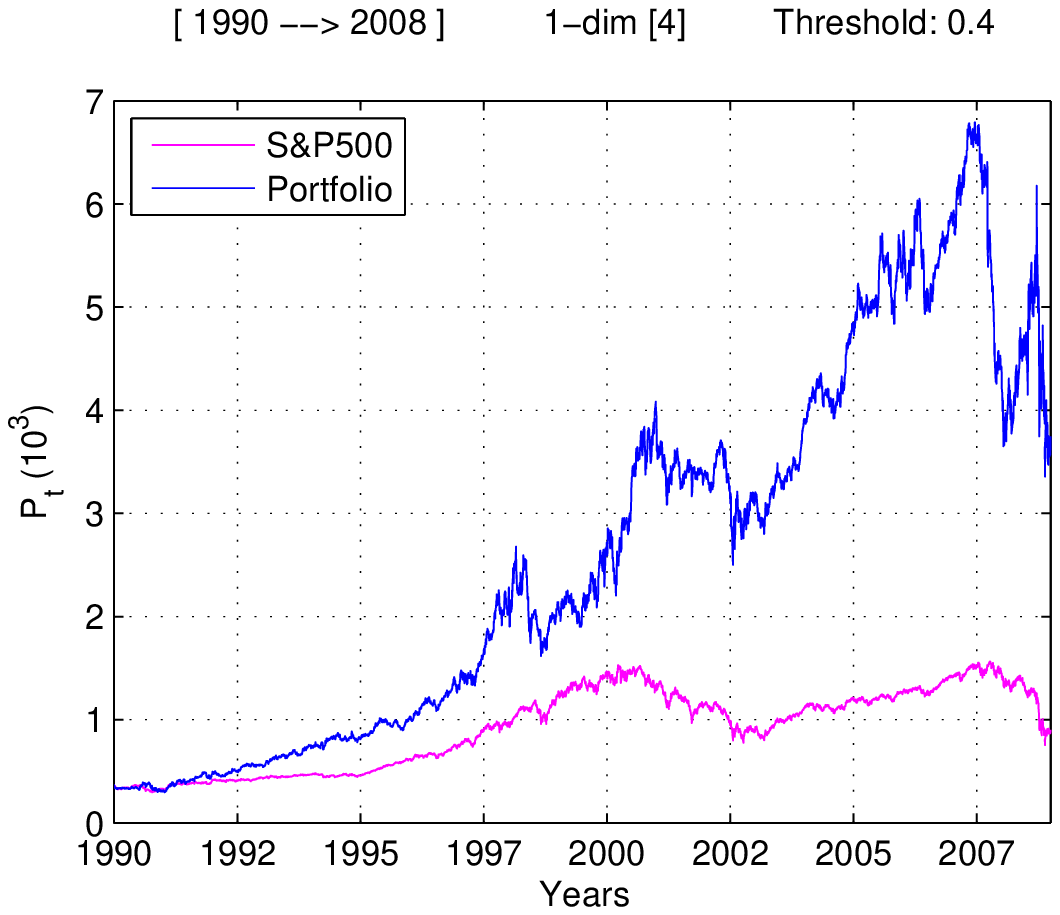}
\caption{Direction-4 portfolio (with threshold 0.4) compared with S\&P500}
\label{dim4}
\end{figure}

\begin{figure}[tbp]
\centering
\includegraphics[width=10cm]{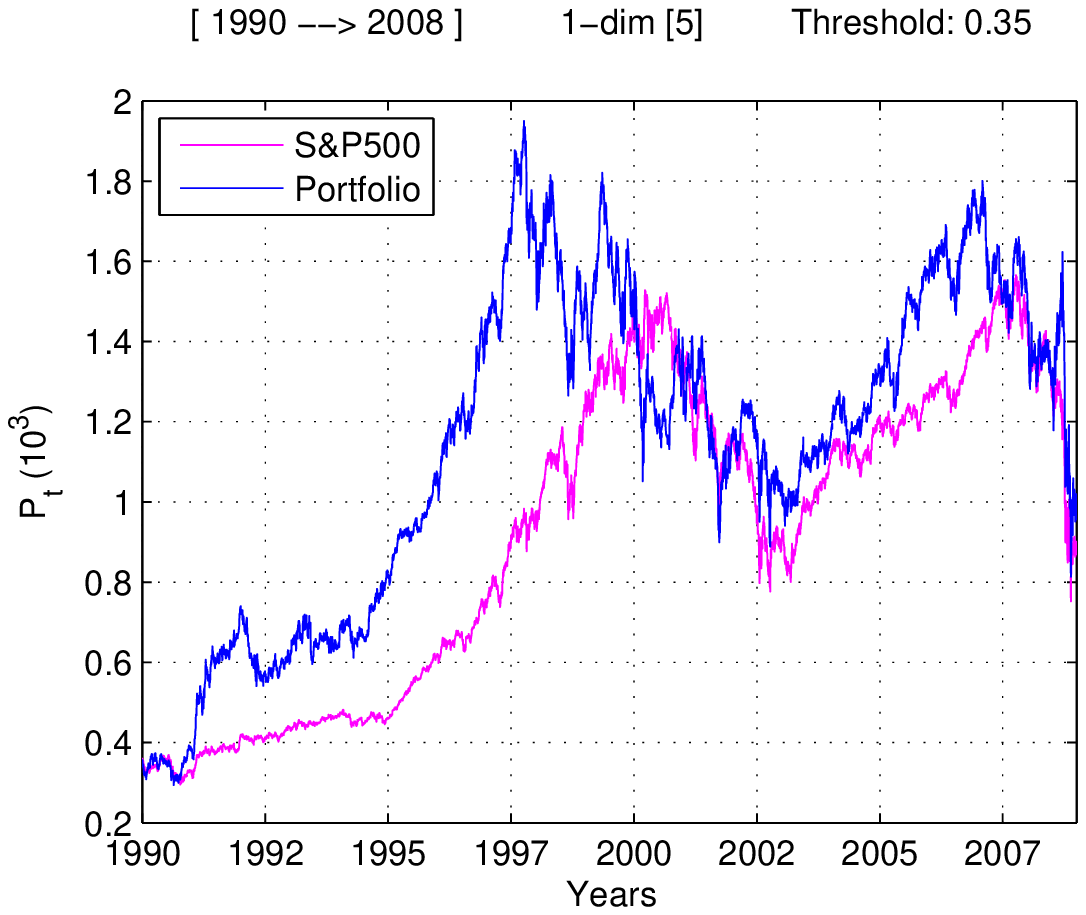}
\caption{Direction-5 portfolio (with threshold 0.35) compared with S\&P500}
\label{dim5}
\end{figure}

\begin{figure}[tbp]
\centering
\includegraphics[width=10cm]{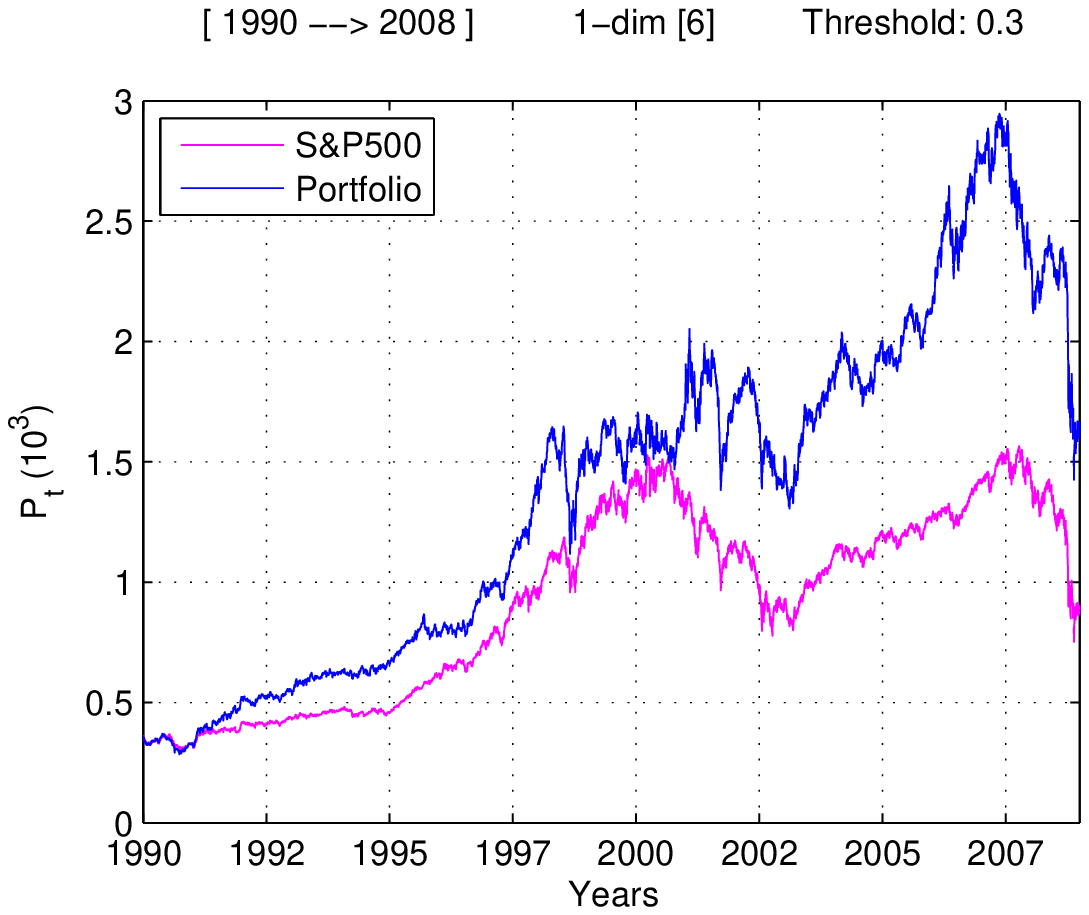}
\caption{Direction-6 portfolio (with threshold 0.3) compared with S\&P500}
\label{dim6}
\end{figure}

Figs.(\ref{dim1_0.4}) to (\ref{dim6}) show the evolution of six
1-dimensional portfolio for each one of characteristic market directions as
well as the simultaneous evolution of the S\&P500 index for a period from
1990 to 2008. These plots were built adjusting each six months the
corresponding portfolios. The average number of companies in each portfolio
ranges from 7 to 42 out of 319. The directions are labeled 1 to 6 from the
largest to the smallest eigenvalue in the market space.

Observing the results in Fig.(\ref{dim1_0.4}) and (\ref{dim1_0.5}) it seems
obvious that building the portfolio on the first direction - the one
associated to the largest eigenvalue - yields negative results when compared
to the evolution of the index. The evolution of a 1-dimensional portfolio
built on the second effective direction remains very close to the value of
the index (Fig.\ref{dim2}) and for the third direction (Fig.\ref{dim3}) two
different behaviors take place. The first one, from 1992 to 2000 yields
negative results whereas from 2004 to 2007 one sees the opposite effect. The
1-dimensional portfolio built on the fifth direction display a similar
behavior (Fig.\ref{dim5}).

However, when the fourth direction is chosen we notice that there is a very
significant gain of 314\% (Fig.\ref{dim4}). Finally, the sixth direction is
also related to gains even though, there seems to be some loses at the end
of the time period (Fig.\ref{dim6}).

In general terms, it is clear that, whenever 1-dimensional portfolios are
built, there are two remarkable directions: the first and the fourth. The
first one is always associated to negative results, while the fourth
direction corresponds to strongly positive ones. Mixed behavior is
associated to the other directions.

\subsection{Higher dimensional portfolios}

\begin{figure}[tbp]
\centering
\includegraphics[width=10cm]{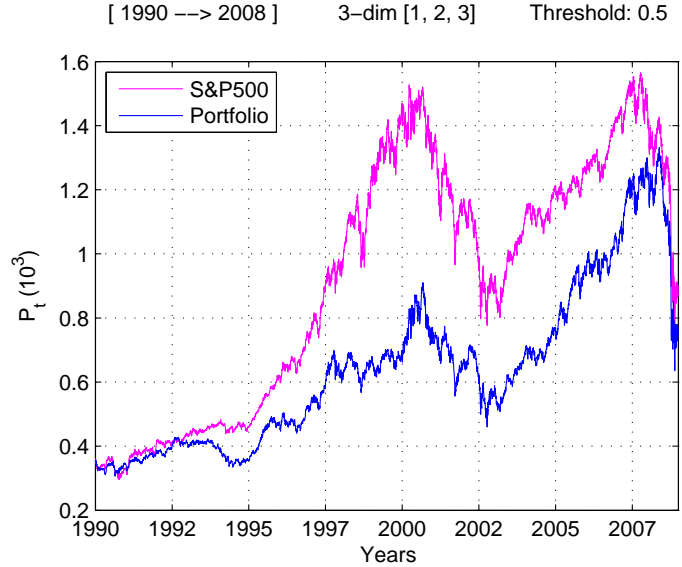}
\caption{Portfolio associated to the 1,2,3-subspace (with threshold 0.5)
compared with S\&P500}
\label{dim123}
\end{figure}

\begin{figure}[tbp]
\centering
\includegraphics[width=10cm]{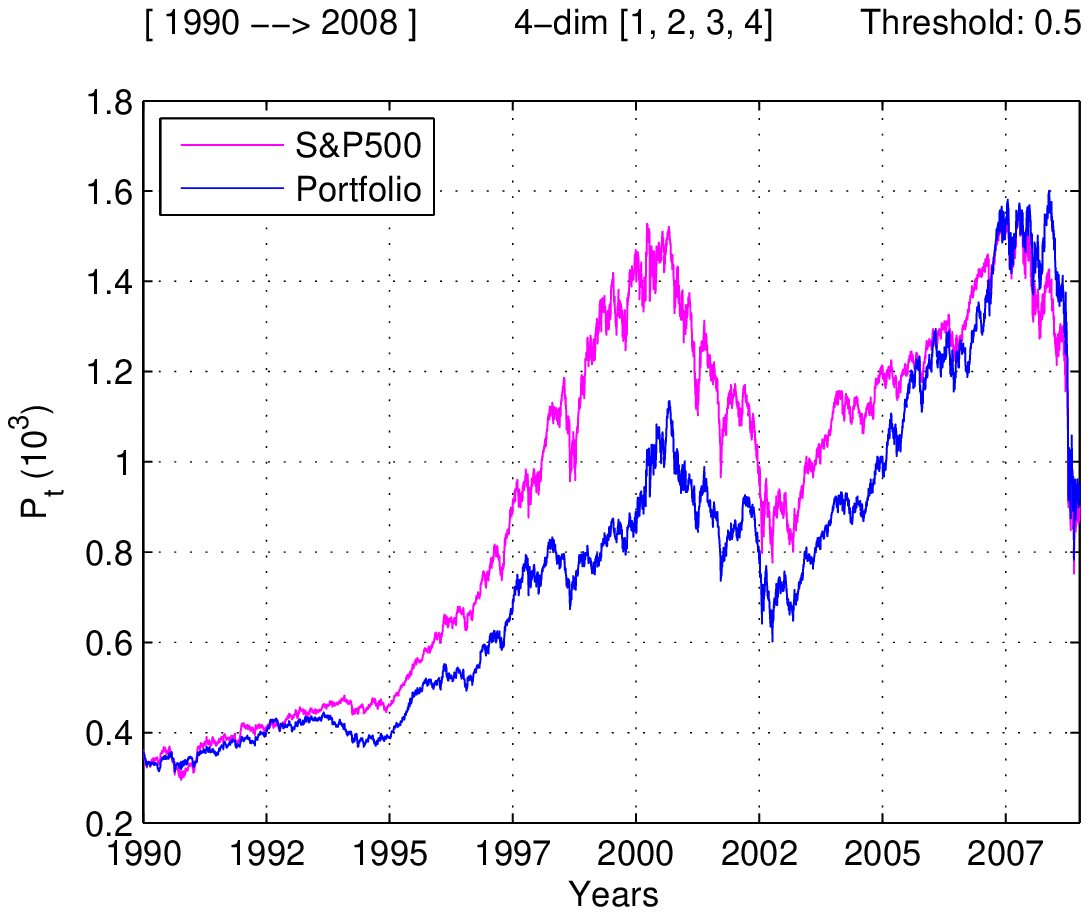}
\caption{Portfolio associated to the 1,2,3,4-subspace (with threshold 0.5)
compared with S\&P500}
\label{dim1234}
\end{figure}
\begin{figure}[tbp]
\centering
\includegraphics[width=10cm]{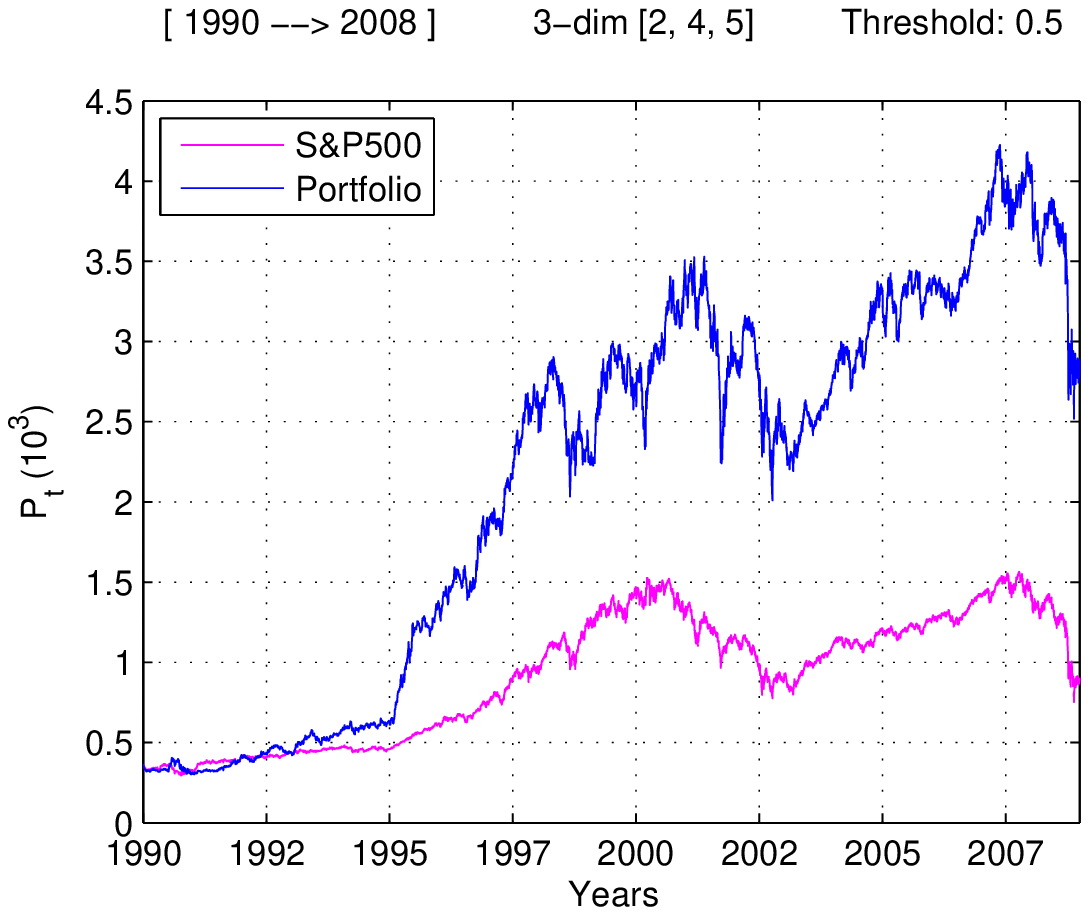}
\caption{Portfolio associated to the 2,4,5-subspace (with threshold 0.5)
compared with S\&P500}
\label{dim245}
\end{figure}

\begin{figure}[tbp]
\centering
\includegraphics[width=10cm]{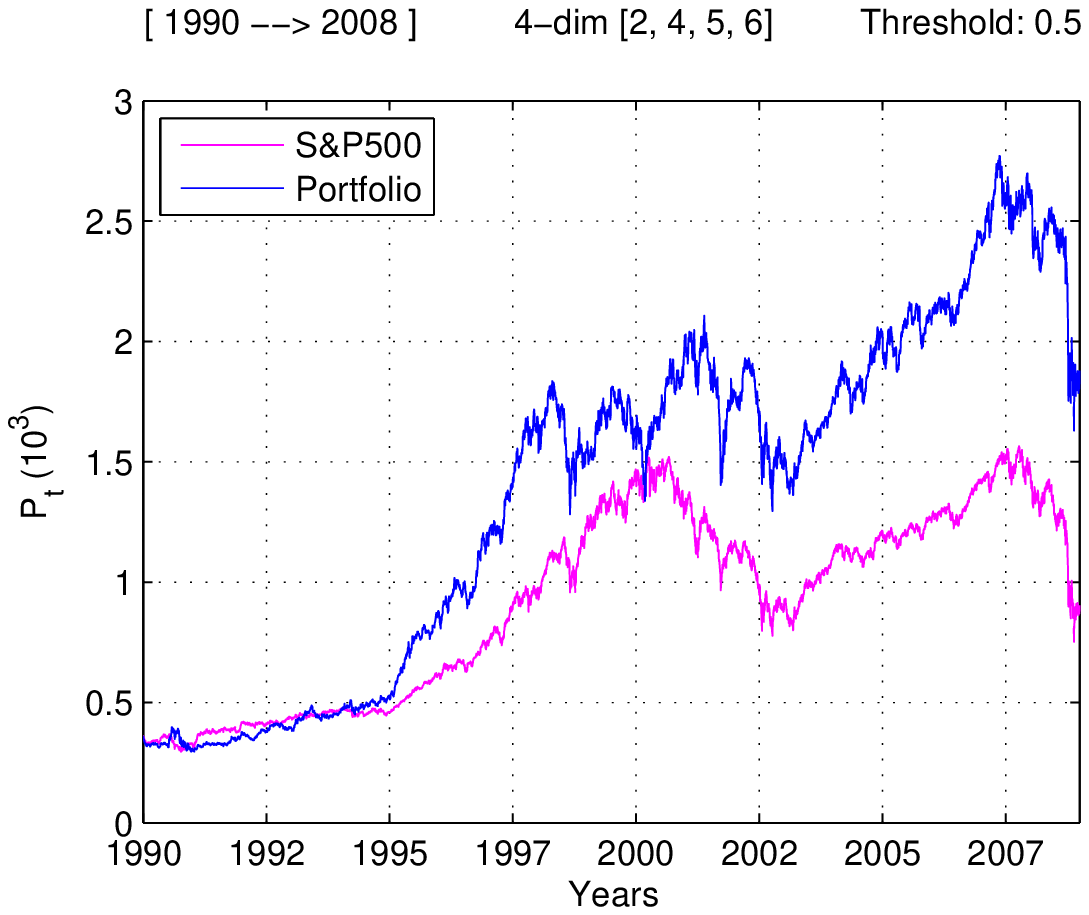}
\caption{Portfolio associated to the 2,4,5,6-subspace (with threshold 0.5)
compared with S\&P500}
\label{dim2456}
\end{figure}

Figs.(\ref{dim123}) to (\ref{dim2456}) show the evolution of
multi-dimensional portfolios compared with the evolution of the S\&P500
index. These plots were also built from a varying number of stocks, adjusted
at each 6-month period from 1990 to 2008.

The negative character of the first dimension is evident even in situations
where more than one dimension is involved, as shown in Figs.(\ref{dim123})
and (\ref{dim1234}) for portfolios built on the subspaces, respectively,
[1,2,3] and [1,2,3,4]. (with threshold $0.5$).

In contrast, 3 and 4-dimensional portfolios containing the second, fourth
and fifth directions seem to have consistently higher values as compared to
the index, as shown in Figs.(\ref{dim245}) and (\ref{dim2456}).

As mentioned before, the most interesting results were obtained from the
fourth direction, for which the gain factor is 314\%. The effect of the
fourth direction extends to the performance of the portfolio built on the
subspace [2,4,5] which yields a gain factor 220\%.

\section{Subspace-portfolios and the Markowitz plane}

Although being questionable that standard deviation of return is a good
measure of risk, the mean-variance method of Markowitz has become a standard
model against which other portfolio constructions may be compared. Here
also, after constructing the efficient frontier for the whole set of
companies that were analyzed, we have computed the mean return and standard
deviation of our portfolios and placed them in the Markowitz plane. 
\begin{figure}[tbp]
\centering 
\includegraphics[width=15cm]{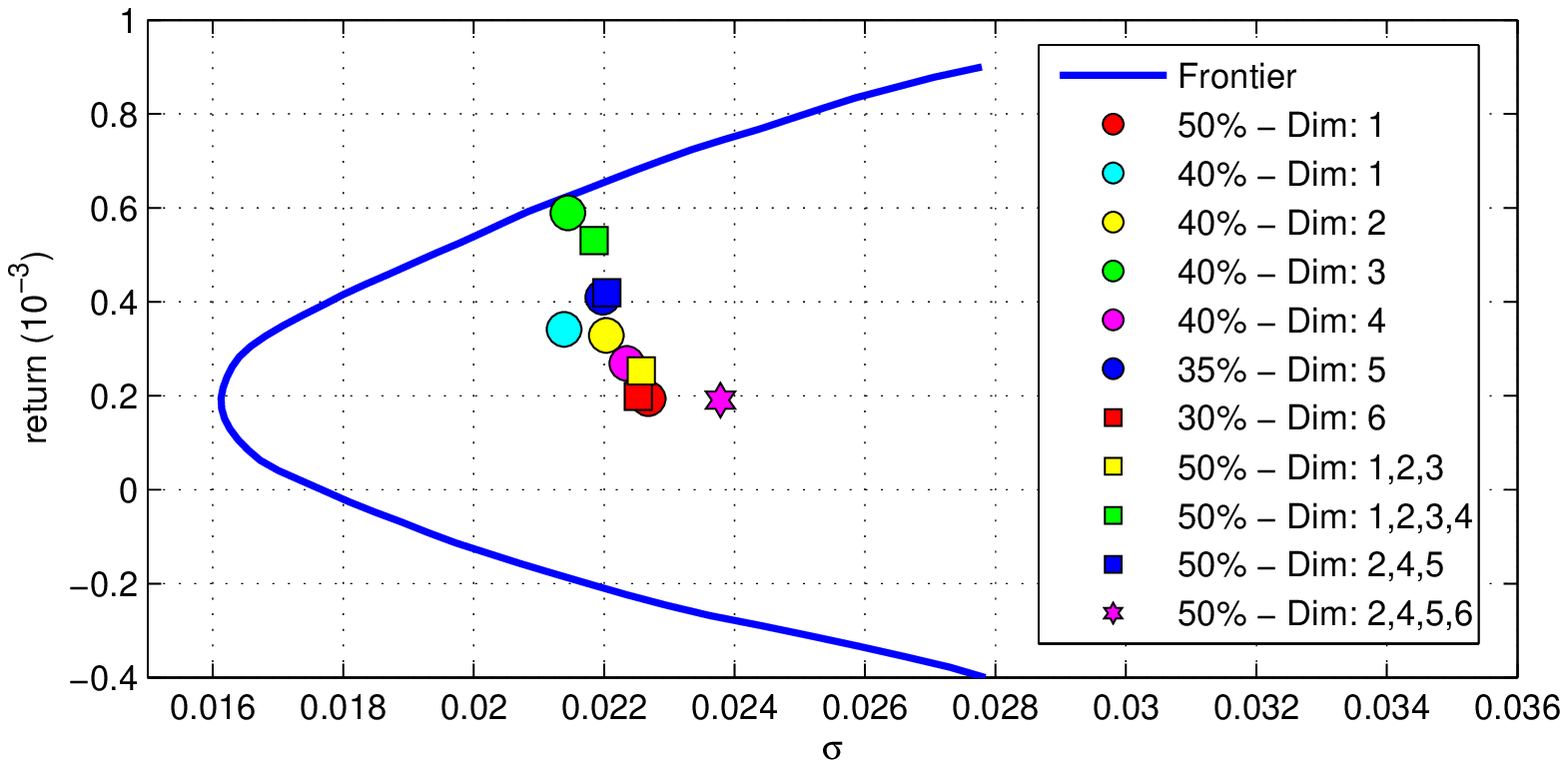}
\caption{The subspace portfolios in the Markowitz plane}
\label{mark_total}
\end{figure}

The efficient frontier is constructed considering portfolios which may
contain positive amounts (no short positions) of any one of the 319
companies that were used in the analysis. Returns and standard deviations
for the whole period from 1990 to 2008 were used. For our portfolios we have
computed the return and standard deviation for each one of the six months
periods and then have taken the average of these results. The results are
shown in Fig.(\ref{mark_total}) with color codes described in the right-hand
table. One finds that the 4th direction portfolio is close to a Markowitz
frontier portfolio as also are two of the portfolios that contain this
direction.

The results presented in the figures refer to average values in the whole
1990-2008 period. The same analysis for each particular six-month interval
displays some fluctuations but it is qualitatively very similar.

\section{Conclusions}

A geometric analysis of the time series of returns has been performed in the
past and it implies that the most of the systematic information of the
market is contained in a space of small dimension. In this paper we have
explored subspaces of this space to find out the relative performance of
portfolios formed from the companies that have larger projections in each
one of the subspaces. The subspace directions are ordered according to the
decreasing values of the eigenvalues in the distances matrix. This analysis
is performed in a dynamical manner, that is, at each six months period the
geometrical analysis is performed anew and the companies contained in each
subspace portfolio may change.

An interesting and, for us, unexpected result was that the best performance
portfolios were associated to some of the small eigenvalue subspaces and not
to the dominant directions in the distances matrix. This occurs in such a
systematic fashion over an extended period (1990-2008) that it may not be a
statistical accident. Whether there is some deep reason or a simple
explanation for this effect is an open question.

\end{document}